\documentclass[tinyml]{acmart}
\usepackage{booktabs}
\usepackage{multirow}
\usepackage{tabularray}
\usepackage{algorithm}
\usepackage{algpseudocode}
\usepackage[table,xcdraw]{xcolor}

\AtBeginDocument{%
  \providecommand\BibTeX{{%
    \normalfont B\kern-0.5em{\scshape i\kern-0.25em b}\kern-0.8em\TeX}}}

\setcopyright{rightsretained}
\copyrightyear{2025}
\acmYear{2025}

\begin{document}

\author{\small Tousif Rahman$^{\dagger*}$ \hspace{0.15em} Gang Mao$^{\dagger*}$ \hspace{0.3em} Bob Pattison$^\dagger$ \hspace{0.3em} Sidharth Maheshwari$^{\ddagger}$ \hspace{0.3em} Marcos Sartori$^{\dagger}$, Adrian Wheeldon$^{\diamond}$, Rishad Shafik$^{\dagger\diamond}$ \hspace{0.15em} Alex Yakovlev$^{\dagger\diamond}$}
\affiliation{%
 \institution{\small $^\dagger$Newcastle University, Newcastle upon Tyne, UK \hspace{0.3em} $^\mathsection$Indian Institute of Technology Jammu, Jammu, India \hspace{0.3em} $^\diamond$Literal Labs, Newcastle upon Tyne, UK}
 \country{}
}
\affiliation{%
  \institution{$^*$Indicates equal contribution. Correspondence: tousif.rahman@newcastle.ac.uk}
  \country{}
}

\renewcommand{\shortauthors}{Rahman and Mao et al.}
\renewcommand{\authors}{Tousif Rahman, Gang Mao, Bob Pattison, Sidharth Maheshwari, Marcos Sartori, Adrian Wheeldon, Rishad Shafik, Alex Yakovlev}
\title{Runtime Tunable Tsetlin Machines for Edge Inference on eFPGAs}

\begin{abstract}
Embedded Field-Programmable Gate Arrays (eFPGAs) allow for the design of hardware accelerators of edge Machine Learning (ML) applications at a lower power budget compared with traditional FPGA platforms. However, the limited eFPGA logic and memory significantly constrain compute capabilities and model size. As such, ML application deployment on eFPGAs is in direct contrast with the most recent FPGA approaches developing architecture-specific implementations and maximizing throughput over resource frugality. This paper focuses on the opposite side of this trade-off: the proposed eFPGA accelerator focuses on minimizing resource usage and allowing flexibility for on-field recalibration over throughput. This allows for runtime changes in model size, architecture, and input data dimensionality without offline resynthesis. This is made possible through the use of a bitwise compressed inference architecture of the Tsetlin Machine (TM) algorithm. TM compute does not require any multiplication operations, being limited to only bitwise \texttt{AND}, \texttt{OR}, \texttt{NOT}, summations and additions. Additionally, TM model compression allows the entire model to fit within the on-chip block RAM of the eFPGA. The paper uses this accelerator to propose a strategy for runtime model tuning in the field. The proposed approach uses 2.5x fewer Look-up-Tables (LUTs) and 3.38x fewer registers than the current most resource-fugal design and achieves up to 129x energy reduction compared with low-power microcontrollers running the same ML application.  
\end{abstract}


\maketitle

\section{Introduction}
\label{Sec:introduction}
Deploying Machine Learning (ML) applications on the edge involves finding an adequate trade-off between the performance of the ML model, in terms of accuracy and the compute and memory constraints of the edge device~\cite{Edge_Review, Verhelst}. 
\begin{figure}[h]
    \centering
    \includegraphics[width = 0.99\linewidth]{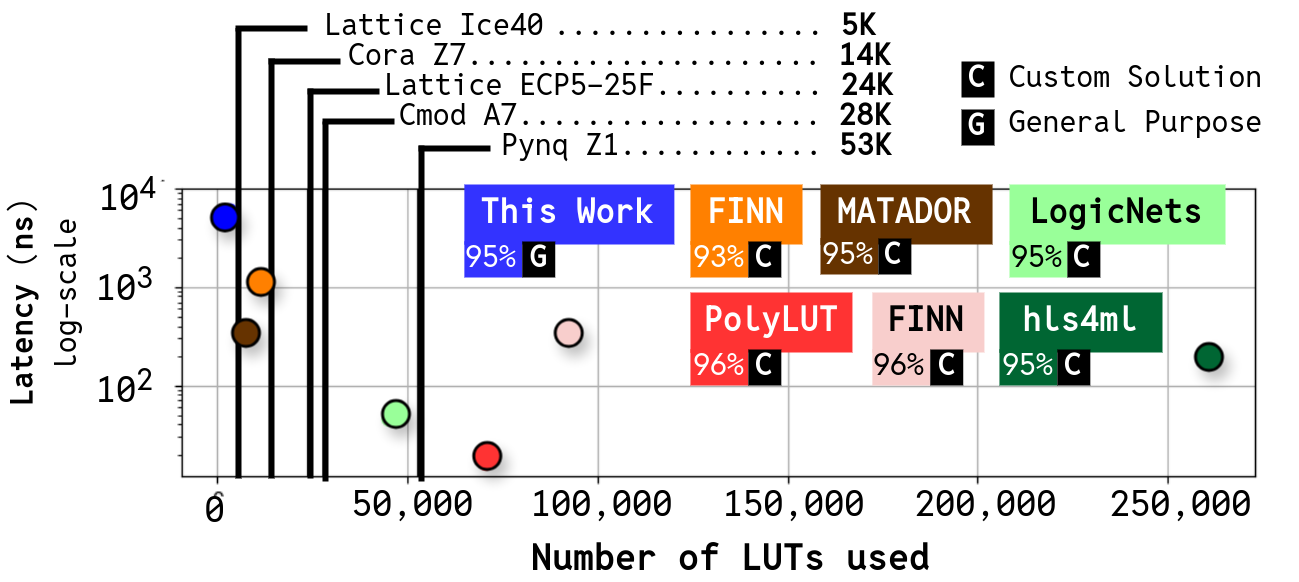}
    \vspace{-3.5mm}
    \caption{\small{Comparing the proposed design (3480 LUTs configuration) to state-of-the-art accelerator automation flows targeting FPGAs. All accelerators were designed for MNIST. Each vertical line indicates the max LUTs of an off-the-shelf eFPGA platform. This work uses \textit{2.5x} fewer LUTs than the next closest work (MATADOR).}} 
    \label{fig:motivation}
    \vspace{-5mm}
\end{figure}
\begin{figure}[h]
    \centering
    \includegraphics[width = 0.99\linewidth]{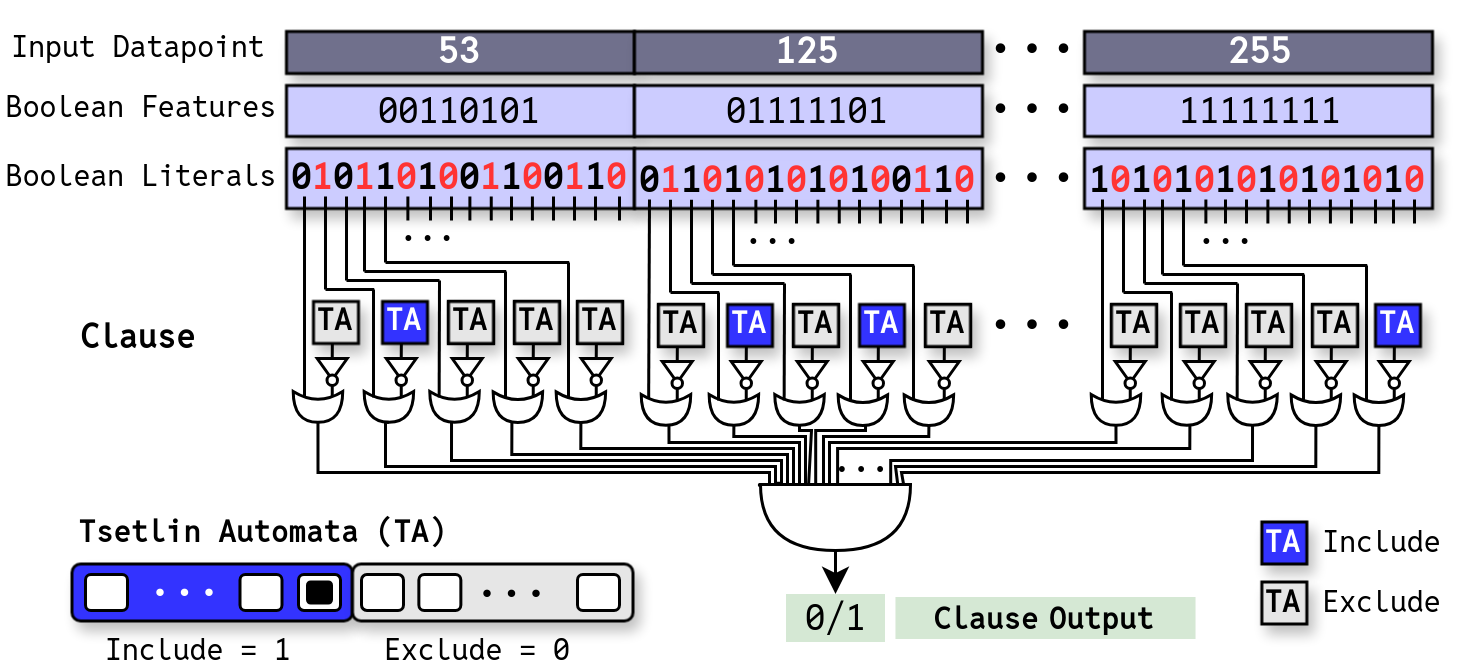}
    \vspace{-3.5mm}
    \caption{\small{Core components of the Tsetlin Machine: Input conversion to Boolean literals, the Tsetlin Automata (TA) and Clause compute.}}
    \label{fig:clause}
    \vspace{-6.5mm}
\end{figure}
\vspace{-0.5mm}The trade-off is made more challenging when considering the compute cost of floating-point multiply-accumulate (MAC) arithmetic and 32-bit model sizes of traditional Deep Neural Networks (DNNs). Fortunately, these challenges can be somewhat alleviated by quantizing DNN weights and data, often to a single bit producing Binary Neural Networks (BNNs)~\cite{FINN_R, polylut, FracBNN, HLS4ML, Conti_2018}. This means that the MAC is simplified to \texttt{XNOR} and \texttt{popcount} operations and weights in memory now use 1-bit each. This approach is particularly effective when translated to \textit{accelerators} targeting Field Programmable Gate Arrays (FPGAs). The most recent works follow one of two strategies: efficiently mapping quantized computation to the FPGA's Look-up-Tables (LUTs) by converting the memory elements into custom compute ~\cite{polylut, logicnets, LUTNet, MATADOR}. Or, they develop architecture-specific approaches for the ML model that are centered around a parameterized compute engine~\cite{FINN_R, FracBNN, SveinConvTM, HLS4ML}. Often both types of approach are also wrapped in design automation flows.

Fig.~\ref{fig:motivation} shows recent state-of-the-art automation flows that can generate accelerators for FPGAs, including this work. The figure shows that the chosen trade-off in current FPGA accelerator work is maximizing inference throughput through custom designs at the expense of LUTs. In doing so, most works are unable to deploy their accelerators onto smaller, cheaper, and more power efficient embedded FPGA (eFPGA) platforms, even for a simple MNIST~\cite{mnist} application (70K LUTs for PolyLUT~\cite{polylut} and 260K LUTs using hls4ml~\cite{HLS4ML}). 

\begin{figure*}[h]
    \centering
    \includegraphics[width =0.96\linewidth]{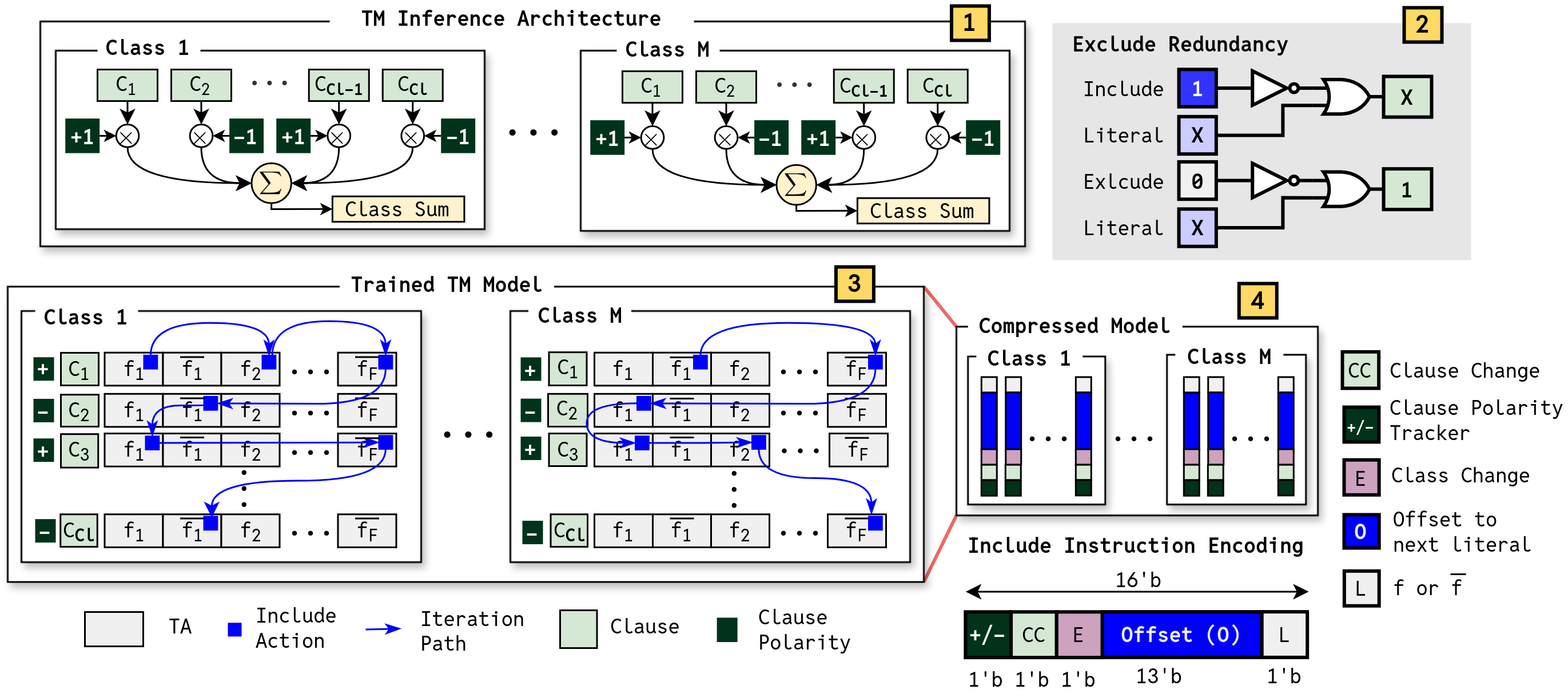}
    \vspace{-2.5mm}
    \caption{\small{1: The class sum compute in the original TM algorithm. 2: The impact of Includes and Excludes in the Clause Output computation - showing that excludes become redundant during inference. 3: The traversal of a trained TM model using when only considering included TAs. 4: The encoding instruction used to create a compressed TM model adapted from the approach used by~\cite{REDRESS}.}} 
    \label{fig:compression}
    \vspace{-5.3mm}
\end{figure*}

This paper focuses on the alternative side of this trade-off: minimizing resource utilization as much as possible to target smaller eFPGA platforms and, crucially, favoring \textit{flexibility} at the cost of longer latency. The proposed accelerator introduces flexibility through \textit{real-time reconfiguration}, which will be referred to as run-time tunability henceforth, of the ML model size and architecture, as well as input data size that allow adaptability to new workloads \textit{without} offline resynthesis to generate a new bitstream and reprogram the eFPGA. This flexibility is made possible by the recent ML algorithm upon which the proposed accelerator is built called the \textit{Tsetlin Machine} (\textit{TM})~\cite{granmo2021tsetlin}. Unlike current DNN quantization approaches, the TM is inherently logic based; it does not require quantization, as the main computation is \textit{already} bitwise (\texttt{AND}, \texttt{OR} and \texttt{NOT}). Fig.~\ref{fig:clause} shows the fundamental computation components of the TM used to generate a \textit{Clause Output}.

\textbf{Boolean Inputs:} Starting at the top of Fig.~\ref{fig:clause}, the input data (Input Datapoint) is converted into \textit{Boolean Literals}. For small edge applications, this is simply the binary representation of the data, referred to as Boolean features, and their complements (seen in red). This process is called \textit{Booleanization}. 

\textbf{Clause Computation: } Each Boolean literal interacts with its own learning element called a \textit{Tsetlin Automata} (\textit{TA}). The TA is a finite state machine with states corresponding to one of two actions: \textit{Include} or \textit{Exclude} producing a \texttt{1} or \texttt{0} respectively. The model training process finds optimum Include/Exclude state for each TA (detailed in~\cite{granmo2021tsetlin,XOR_TPAMI, olga}). The actions of each TA are inverted and \texttt{OR}'ed with their Boolean literal, they are all then \texttt{AND}'ed to form the 1-bit \textit{Clause Output} (seen in green).  

\textbf{TM Inference Architecture: } Fig~\ref{fig:compression}.\textbf{1} shows the architecture of the TM. In a multi-class classification problem with \texttt{M} classes, each class has \texttt{Cl} Clauses where each clause will have its own set of TAs, as seen in Fig.~\ref{fig:clause}, and generate a 1-bit Clause output. Clauses in each class have a polarity +1/-1 (seen in dark green). Each clause output is multiplied by its respective polarity and summed to generate the class sum for each class. The argmax of these class sums produces the predicted class. The size and architecture of the TM is controlled by the number of Boolean literals, the number of clauses, and the number of classes. For example, in keeping with the MNIST example, if MNIST has 784 Boolean features, it will have 1568 Boolean literals. If there are 200 clauses per class (10 classes in MNIST) then altogether this TM model will have 3,136,000 TAs. This paper exploits the compute and structural simplicity of the TM, along with the sparsity and redundancy of the model, to make the following \textbf{contributions}:
\begin{itemize}
    \item \textbf{Implementation:} An LUT frugal, real-time architecture adaptable accelerator. This functionality alone differentiates the proposed design from current FPGA works. In addition, customization of the design is offered if supported by the eFPGA's resources.  
    \item \textbf{Runtime Tunability:} An implementation flow for all proposed accelerator design configurations and a strategy to periodically update the accelerator in real-time without resynthesis once deployed if re-calibration is required.
\end{itemize}
The rest of the paper is organized as follows: Section II focuses on the sparsity and compression opportunities in the TM algorithm. Section III develops these ideas into the proposed accelerator design and its different configurations, as well as the flow for generating the implementation and its on-field runtime tunability. Section IV benchmarks the accelerator against comparable work, and Section V concludes the paper.

\begin{figure*}[h]
    \centering
    \includegraphics[width =\linewidth]{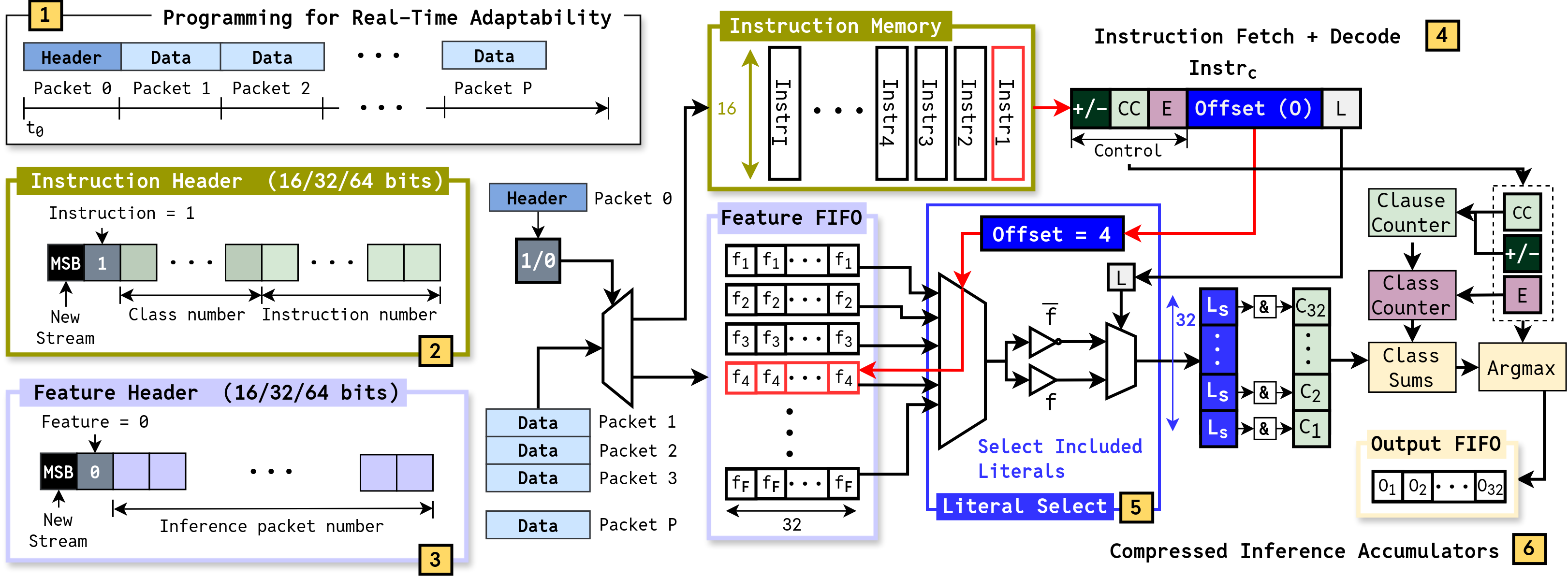}
    \vspace{-4mm}
    \caption{\small{Overview of the Proposed Accelerator (Base Version): 1: An incoming data stream to the accelerator; the header packet of the stream is used for configuration. 2: The bit-fields of the header when the data stream contains instructions (the TM model) 3: The bit-field of the header when the data stream contains input Boolean features. 4: The instruction fetching and decoding process. 5: Selecting Boolean literals that match the TA Include actions. 6: Accumulators for the clause outputs and class sums when performing the compressed inference.}} 
    \label{fig:arch}
    \vspace{-5mm}
\end{figure*}

\section{Sparse Tsetlin Machines}
For inference, the state inside each TA in a TM model is represented only by a 1-bit Include or Exclude action (see Fig.~\ref{fig:clause} if needed). During TM training, this leads to a very sparse model where the number of Excludes vastly outnumbers the Includes~\cite{Bakar2023,REDRESS, litbudget}. This sparsity can be exploited when examining Fig~\ref{fig:compression}.2. It shows the impact of an Include and an Exclude on the clause output. The Exclude eliminates the contribution of an input Boolean literal (\texttt{X} $\in$ \{0,1\}). However, an Include propagates the Boolean literal. This means \textit{only TAs that are Includes are needed for TM inference}.

\textbf{Compression: } The accelerator design builds on this Include-only approach by adapting a compression scheme similar to~\cite{REDRESS} which claims around 99\% model compression for small edge datasets. Continuing with the MNIST example, if there are 3,136,000 TA actions in total, only around 17,000 will be Include actions. A visual overview of this approach can be seen in Fig.~\ref{fig:compression}.3 showing the structure of a trained TM model. Within each class, in each row are the TA actions for each Boolean literal that makes up a clause (These TA actions are written as \texttt{f} and $\overline{\texttt{f}}$ to signify the Boolean feature and its complement that the TA action corresponds to). Therefore, each row represents a clause output, and class sums can be formed using each clause's corresponding polarity (+/-).

Continuing with Fig.~\ref{fig:compression}.3, a little blue box indicates where a TA action is an Include. The model can therefore be traversed iteratively using \textit{ only} these TAs for inference. This is shown with the blue arrow that indicates the iteration path over all the TA Include actions for each class. In most TM applications, the TAs with Include actions make up around 1\% of the total model. Therefore, this iteration path is much shorter than computing every TA for every clause in every class in the model.  

\textbf{Include Instruction Encoding: }The compressed model only needs these included actions. However, each TA Include action (little blue box) now needs to be encoded with information about its respective Boolean literal, clause, clause polarity and class. This is achieved through a 16-bit encoding called the Include Instruction Encoding (Fig~\ref{fig:compression}.4). This instruction encoding contains the necessary information to jump from one TA Include action to another so the inference can be done directly using this compressed model. The jump is controlled with an offset bit field, \texttt{Offset(O)} (seen in blue), indicating the number of TAs until the next TA Include action. For each action, the encoding toggles the two MSB bits (\texttt{+/-}, \texttt{CC}) each time a clause changes, and the LSB (\texttt{L}) indicates whether the Boolean Literal is the feature \texttt{f} or its complement $\overline{\texttt{f}}$. This paper also adds an extra bit \texttt{E} that toggles when changing classes. Compressed inference forms the backbone of the proposed accelerator architecture and is used by other authors dealing with training TMs to become more sparse~\cite{litbudget} or exploiting this sparsity for TM inference - in particular targeting micro-controllers (MCU)~\cite{Bakar2023, REDRESS} claiming substantial speed-ups (up to 5700x) compared to embedded BNNs. 

\vspace{-3.6mm}
\section{Design and Runtime Tunability}
\label{Sec:arch}
The proposed architecture performs TM inference using the compressed instructions described in the previous section. Through Fig.~\ref{fig:arch} the main attributes of the accelerator design will be highlighted: Real-time adaptability of the model and task, customization options, and resource frugality:

\textbf{Programming for Real-time Adaptability:} The accelerator can be reconfigured for different TM models and input data sizes in real-time using a data stream as seen in Fig.~\ref{fig:arch}.1. The data packets start with a header to configure the architecture. The header is either an Instruction Header (Fig.~\ref{fig:arch}.2) to send a new TM model, or a Feature Header (Fig.~\ref{fig:arch}.3) to send Boolean features for inference.

\textbf{Headers} can be configured as 16,32 or 64-bits. The MSB bit indicates that this is a new stream and resets the accelerator. The second bit from the MSB bit indicates whether the following data packets will contain either the Include instructions (Instruction Header), or Boolean Features for inference (Feature Header). 

\begin{figure*}[h]
    \centering
    \includegraphics[width =0.99 \linewidth]{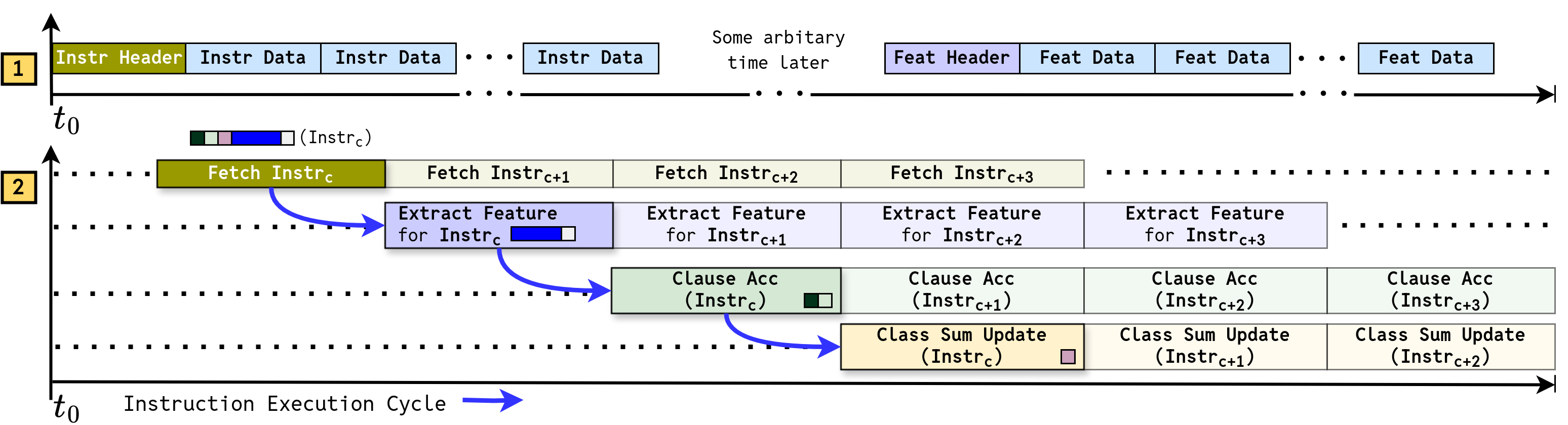}
    \vspace{-4mm}
    \caption{\small{Timing diagrams of the programming, inference and execution cycle of an instruction (1). The instruction execution cycle (2) is a per-core process - it would be the same for a multi-core version of the accelerator.}}
    \label{fig:timing}
    \vspace{-4.8mm}
\end{figure*}

The remainder of the Instruction Header contains the model architecture parameters: the number of classes and the number of clauses. This is used later in the accumulation counters (Clause Counter and Class Counter). The remainder of the Feature header contains the number of Inference data packets the accelerator will need to process to generate classifications. \textbf{Real-time architecture change} is made possible due to the simplicity of the TM structure. Only three parameters are sufficient to update the accelerator to a new TM model size (instruction number in Instruction Header) or a new task and new data dimensionality (Class number in Instruction Header and Inference packet number in Data Header). 

\begin{figure}[h]
    \centering
    \includegraphics[width =0.98\linewidth]{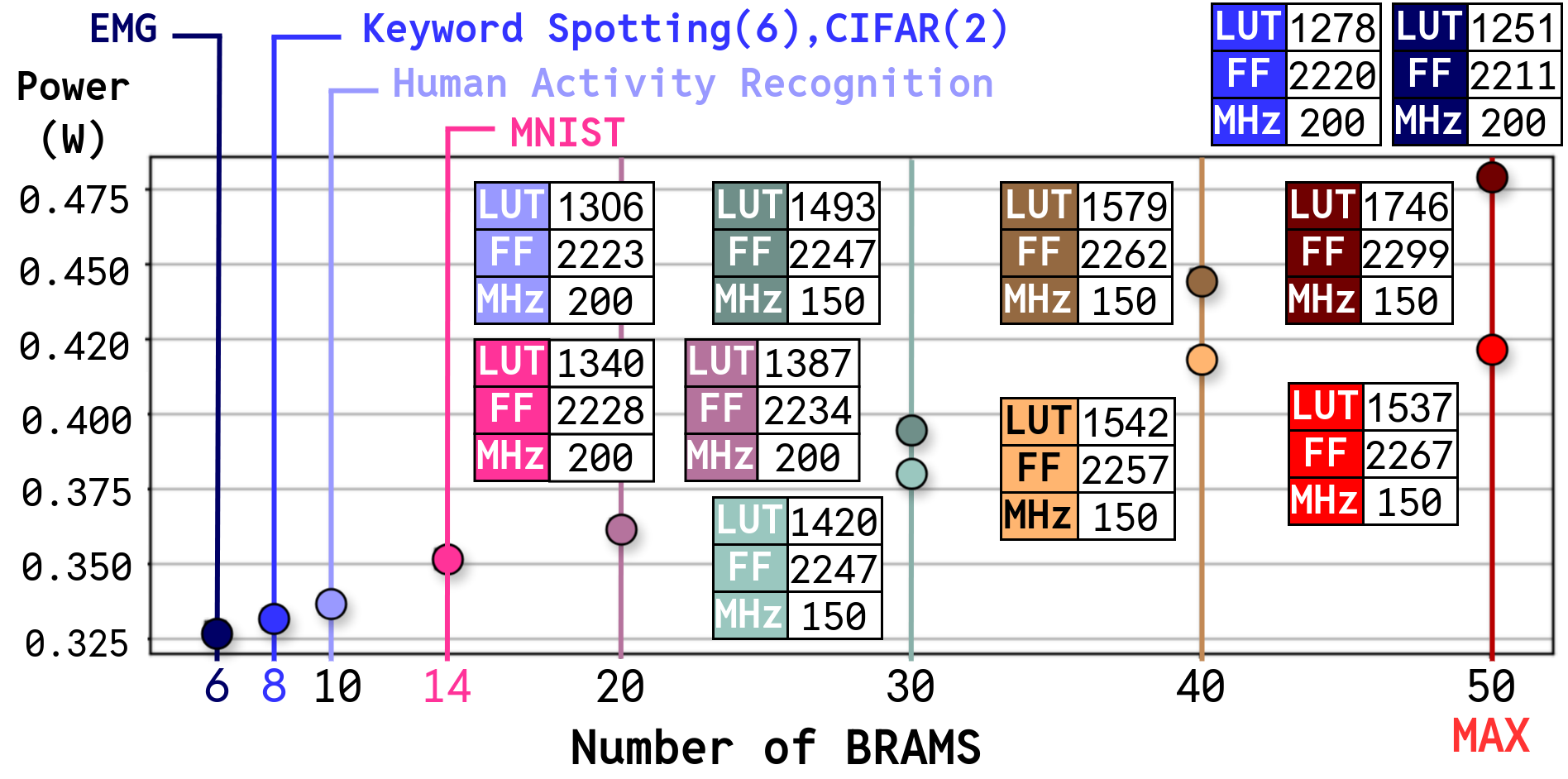}
    \vspace{-2mm}
    \caption{\small{Customization options for memory depths for the base configuration (implemented on Artix A7-35T).}} 
    \vspace{-2.4mm}
    \label{fig:memory}
    
\end{figure}

\textbf{Memory Customization Options:} While the accelerator is general-purpose, the reconfigurability of the eFPGA allows for some customization options when it comes to using memory. So far, the paper has only discussed LUT frugality; however, now it is important to introduce \textit{Block RAM (BRAM) frugality}. The level of compression is possible through the Include-only approach allowing nearly all TM models for edge applications to fit \textit{well} within the BRAM of the smallest Xilinx chips. The eFPGA gives users freedom to configure memory depths for greater runtime tunability options when deployed. However, this comes at the expense of more LUT, Flip Flop (FF), and power usage and running at a lower frequency. However, it allows users to have greater customization than an equivalent fixed-memory ASIC. This is seen through Fig.~\ref{fig:memory} where vertical lines represent the minimum memory required for edge-scale datasets for which this accelerator design is appropriate (discussed in more detail in Section IV). 

These datasets give an indication of the scale of problems this accelerator will be expected to be used for, for example, multivariate sensor data, wearable and telemetry data (e.g. activity detection) and small audio (e.g. keyword detection) and computer vision classification problems (e.g. object classification). These problems do not place as high a priority on \textit{throughput} - unlike the datasets that PolyLUT and LogicNets have targeted like UNSW-NB15~\cite{Network_Intrusion}.

\textbf{Compressed Inference and batching support:} Once the instructions have been loaded into memory, the accelerator can begin the compressed inference computation. This starts by fetching an instruction from the Instruction Memory and ``decoding'' it (Fig.~\ref{fig:arch}.4). The instruction will contain the offset to select the appropriate Boolean feature(s) (if in batched mode) from the Feature Memory. Notice in Fig.~\ref{fig:arch}.5, the \textbf{Literal Select} box, the Offset is \texttt{4} and the 4th element in the Feature Memory is selected. The \texttt{L} bit is then used to identify which Boolean literal is included from this Boolean feature (either \texttt{f} or $\overline{\texttt{f}}$). The next step is to evaluate the clause output. The selected literal (in the blue) is \texttt{AND}'ed with the clause output registers. Notice that there are 32 of the same literal (\texttt{L}$_s$), this is due to the batching support of the accelerator, and 32 datapoints can be computed at once (the same literal but for 32 datapoints). 

\begin{figure}[h]
    \centering
    \includegraphics[width = 0.88\linewidth]{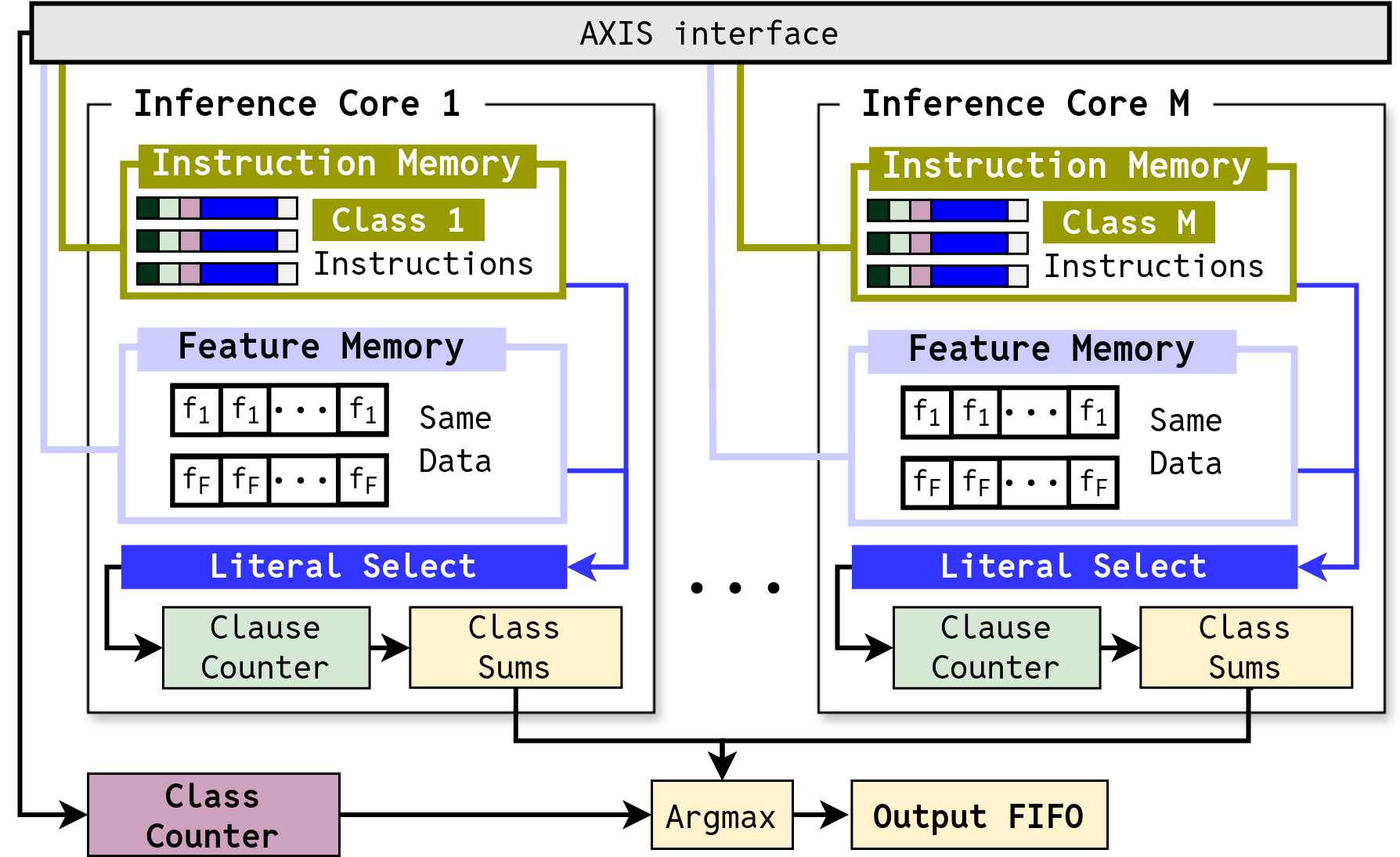}
    \caption{\small{Block diagram for the multi-core design of the accelerator. Each Inference Core is the base version seen in Fig. 4.}} 
    \label{fig:multi_thread_redress}
    \vspace{-6.5mm}
\end{figure}

This clause output continues to be updated by \texttt{AND}ing with included Boolean literals until \texttt{CC} and \texttt{+/-} toggle to say that the clause is completed. The clause output and its polarity are then added to the class sum. The \texttt{E} bit toggles for when classes change. In this way, the full TM model can be iterated in the compressed domain. After iterating through all classes, the argmax is taken from the class sums, and the output FIFO is filled with up to 32 classifications (if in batched mode). This process can be seen through the timing diagram of the instruction execution cycle in Fig.~\ref{fig:timing}. Fig.~\ref{fig:timing} shows the relevant parts of the instructions that are used by each computation stage and the pipelining of this execution cycle. Each instruction takes a minimum of four clock cycles to execute.

\definecolor{Periwinkle}{rgb}{0.8,0.8,1}
\definecolor{SurfCrest}{rgb}{0.835,0.909,0.831}
\definecolor{BarleyWhite}{rgb}{1,0.949,0.8}
\begin{table}
\centering
\scriptsize
\caption{\small{Resource usage of the three proposed accelerator configurations against MATADOR (MTDR) for CIFAR, KWS and MNIST}}
\begin{tblr}{
  width = \linewidth,
  colspec = {Q[280]Q[125]Q[156]Q[171]Q[120]Q[90]},
  row{2} = {Periwinkle},
  row{3} = {SurfCrest},
  row{4} = {BarleyWhite},
  vlines,
  hline{1,8} = {-}{0.08em},
  hline{2,5-7} = {-}{},
}
\textbf{Accelerator Configurations } & \textbf{eFPGA chip} & \textbf{No. LUTs} (LUT-6)        & \textbf{No. FFs}          & \textbf{No. BRAMs} & \textbf{Freq} (MHz) \\
Base (\textbf{B})                     & A7035             & \textbf{1340}       & \textbf{2228}        & 14             & \textbf{200}                 \\
Single Core (\textbf{S})              & Z7020             & 3480 & 5154  & 43         & 100                 \\
Multi-Core (\textbf{M})               & Z7020             & 9814          & 10909 & 43         & 100                 \\
MTDR (CIFAR)              & Z7020             & 3867                & 33212                & \textbf{3}     & 50                  \\
MTDR (KWS)               & Z7021             & 6063                & 10658                & \textbf{3}     & 50                  \\
MTDR (MNIST)              & Z7020             & 8709                & 17440                & \textbf{3}     & 50                  
\end{tblr}
\vspace{-7.3mm}
\label{tab:setup}
\end{table}

\textbf{Configurations (Standalone, Single Core and Multi-Core:} Building on the memory customization options possible with eFPGAs, it is also possible to implement different configurations of the base architecture presented in Fig.~\ref{fig:arch}. Three configurations are offered, the standalone accelerator, an AXI-Stream (AXIS) interfaced single core, and an AXIS connected multi-core. The multi-core architecture is presented in Fig.~\ref{fig:multi_thread_redress}. The AXIS interfacing allows the use of a processor for pre-processing if this is not already done by the edge sensor. Developers may choose the configuration most suitable for their application and ML pipeline. The multi-core design instances base inference cores, but each instruction memory is now loaded with instructions corresponding to non-overlapped classes but the same features into feature memory. The AXIS interface will split the instruction stream and write them into different cores. This allows for class-level parallelism and reduced latency at the cost of more resource usage.

\begin{figure}[h]
    \centering
    \includegraphics[width =0.86\linewidth]{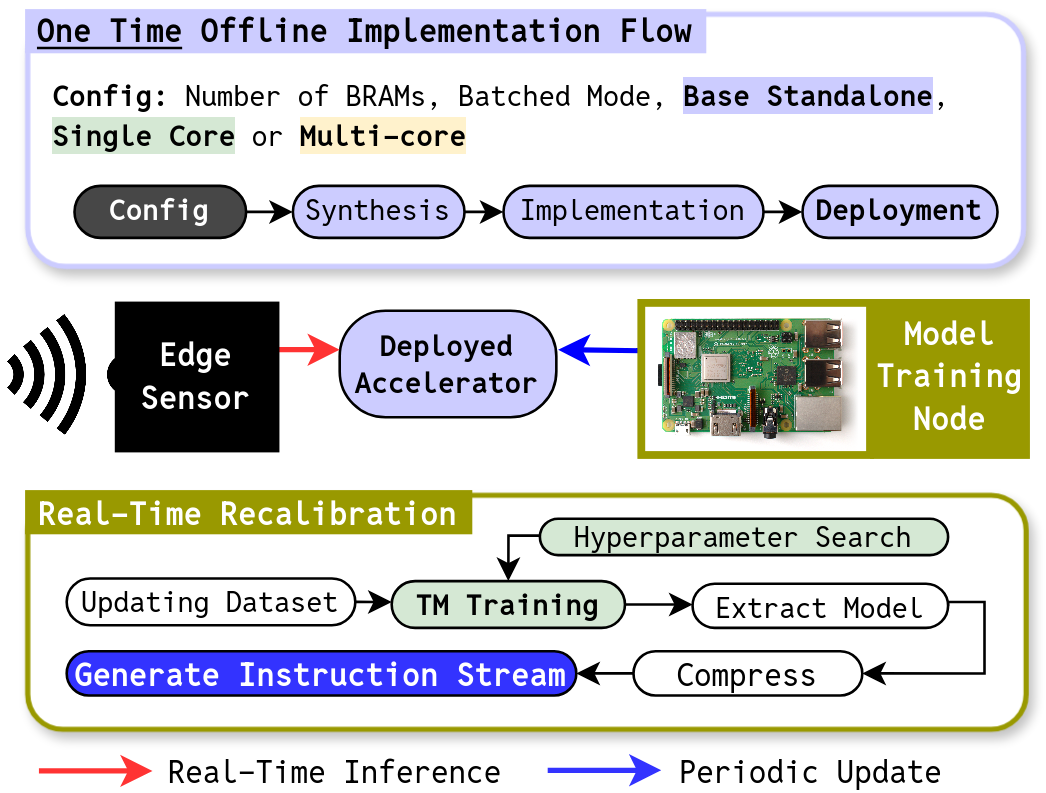}
    \caption{\small{Configuration options for the initial deployment and the proposed system for on-field re-calibration and task update.}} 
    \label{fig:recal}
    \vspace{-4.mm}
\end{figure}

\textbf{Runtime tunability:} One of the main functionalities of the architecture is its flexibility to change the model and the task. There may be instances where data set on which the model was trained no longer adequately reflects the actual data (edge sensor readings may vary subject to aging, temperature, humidity, etc...~\cite{Concept_Drift}). Fig.~\ref{fig:recal} shows a potential system for real-time recalibration. First, a one-time implementation of the accelerator is required. developers can customize memory, batch mode, number of cores and whether it is standalone or AXIS interfaced. Once deployed, the accelerator performs real-time inference from edge sensor data. However, on the same local network (or directly connected) is a Model Training Node. The simplicity of the TM training algorithm leads to fast convergence and energy-efficient training implementations~\cite{olga, TM_KWS}. The authors in~\cite{TM_KWS} demonstrate that Tsetlin machines can be trained very well on small compute nodes like Raspberry Pis. Therefore, this type of node may train on an updating dataset and periodically reprogram the accelerator with a new model if needed. Users can also run a hyperparameter search to update the architecture if needed, or even add an additional class to the classification task. The TM only has two hyperparameters for training and the authors in~\cite{olga} have highlighted the reduced complexity in the TM architecture search space compared with DNNs. \textbf{The key advantage} is this Raspberry Pi node does not require FPGA synthesis tools to reconfigure the proposed accelerator to a new model or task - unlike current architecture-specific FPGA approaches~\cite{FracBNN, logicnets, polylut, MATADOR, FINN_R}.  

\section{Evaluation}
\label{Sec:eval}
To evaluate the proposed architecture, there are two critical questions to consider: \textbf{Question 1:} How significant is the sacrifice in latency and energy of the proposed architecture compared to its closest comparable custom FPGA implementation? \textbf{Question 2:} How much better is the recalibration approach on an \textit{eFPGA} compared to if the same compressed TA-Include only algorithm was developed as software for the processors of low-power off-the-shelf micro-controllers? 

\begin{figure}[h]
    \centering
    \includegraphics[width =0.95\linewidth]{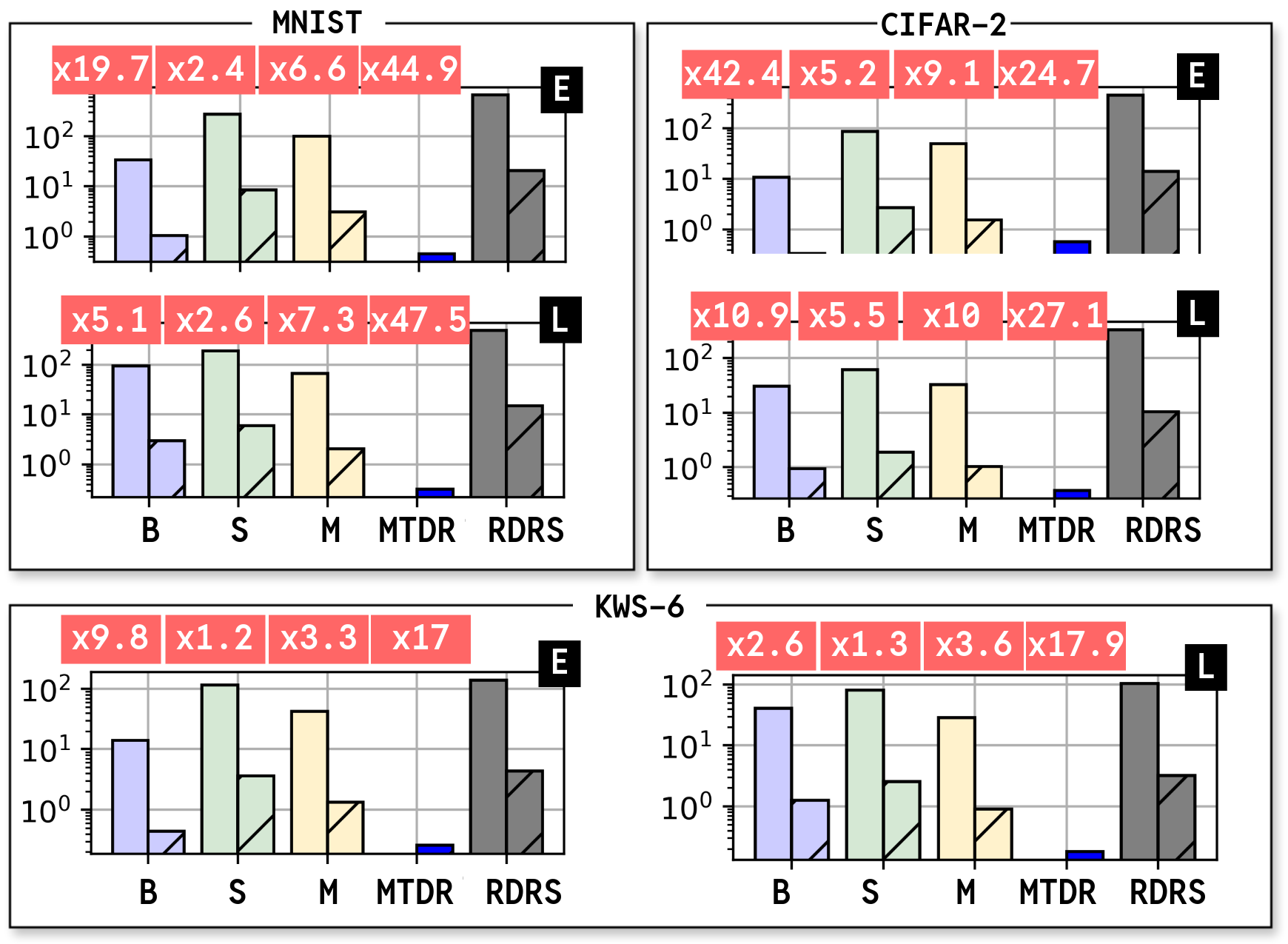}
    \caption{\small{Energy (E) and latency (L) of the proposed accelerator designs (B, S, M) against MATADOR (MTDR) and the same compressed instruction algorithm on the STM32Disco MCU (RDRS). Single datapoint energy and latencies are shown with the \textit{hatched} bar and batched energy and latency are shown with the \textit{solid} bar. MATADOR does not support batch mode, only single datapoint data is reported.}} 
    \label{fig:graph}
    \vspace{-5mm}
\end{figure}

\begin{table*}[]
\centering
\caption{Latency and energy comparisons of the proposed accelerators vs an Espressif ESP32 software version of the algorithm.}
\scalebox{0.90}{
\begin{tabular}{|c|c|c|cc|c|cc|c|c|}
\hline
                                                                                          &                        &                                         & \multicolumn{2}{c|}{Latency (us)}                                                                    &                                                                                & \multicolumn{2}{c|}{Energy (uJ)}                                                                      &                               &                                                                               \\ \cline{4-5} \cline{7-8}
\multirow{-2}{*}{Dataset}                                                                 & \multirow{-2}{*}{Acc}  & \multirow{-2}{*}{Design}                & \multicolumn{1}{c|}{Batch}                                   & Single Data Point                     & \multirow{-2}{*}{\begin{tabular}[c]{@{}c@{}}Throughput\\ (inf/s)\end{tabular}} & \multicolumn{1}{c|}{Batch}                                   & Single data point                      & \multirow{-2}{*}{xSpeedups}   & \multirow{-2}{*}{\begin{tabular}[c]{@{}c@{}}xEnergy\\ Reduction\end{tabular}} \\ \hline
                                                                                          &                        & \cellcolor[HTML]{CCCCFF}Base (B)        & \multicolumn{1}{c|}{\cellcolor[HTML]{CCCCFF}\textbf{7.44}}   & \cellcolor[HTML]{CCCCFF}\textbf{0.23} & \cellcolor[HTML]{CCCCFF}\textbf{4303968}                                       & \multicolumn{1}{c|}{\cellcolor[HTML]{CCCCFF}\textbf{2.610}}  & \cellcolor[HTML]{CCCCFF}\textbf{0.082} & \cellcolor[HTML]{CCCCFF}245.3 & \cellcolor[HTML]{CCCCFF}22.9                                                  \\
                                                                                          &                        & \cellcolor[HTML]{D5E8D4}Single Core (S) & \multicolumn{1}{c|}{\cellcolor[HTML]{D5E8D4}14.87}           & \cellcolor[HTML]{D5E8D4}0.46          & \cellcolor[HTML]{D5E8D4}2151984                                                & \multicolumn{1}{c|}{\cellcolor[HTML]{D5E8D4}21.279}          & \cellcolor[HTML]{D5E8D4}0.665          & \cellcolor[HTML]{D5E8D4}122.7 & \cellcolor[HTML]{D5E8D4}2.8                                                   \\
                                                                                          &                        & \cellcolor[HTML]{FFF2CC}5-Core (M)      & \multicolumn{1}{c|}{\cellcolor[HTML]{FFF2CC}7.64}            & \cellcolor[HTML]{FFF2CC}0.24          & \cellcolor[HTML]{FFF2CC}4188482                                                & \multicolumn{1}{c|}{\cellcolor[HTML]{FFF2CC}11.429}          & \cellcolor[HTML]{FFF2CC}0.357          & \cellcolor[HTML]{FFF2CC}238.7 & \cellcolor[HTML]{FFF2CC}5.2                                                   \\ \cline{3-10} 
\multirow{-4}{*}{\begin{tabular}[c]{@{}c@{}}EMG\\ ~\cite{emg_data_for_gestures_481}\end{tabular}}                      & \multirow{-4}{*}{87\%} & ESP32                                   & \multicolumn{1}{c|}{1824.00}                                 & 57.00                                 & 17544                                                                          & \multicolumn{1}{c|}{59.791}                                  & 1.868                                  & -                             & -                                                                             \\ \hline
                                                                                          &                        & \cellcolor[HTML]{CCCCFF}Base (B)        & \multicolumn{1}{c|}{\cellcolor[HTML]{CCCCFF}37.80}           & \cellcolor[HTML]{CCCCFF}1.18          & \cellcolor[HTML]{CCCCFF}846561                                                 & \multicolumn{1}{c|}{\cellcolor[HTML]{CCCCFF}\textbf{13.268}} & \cellcolor[HTML]{CCCCFF}\textbf{0.415} & \cellcolor[HTML]{CCCCFF}490.2 & \cellcolor[HTML]{CCCCFF}109.4                                                 \\
                                                                                          &                        & \cellcolor[HTML]{D5E8D4}Single Core (S) & \multicolumn{1}{c|}{\cellcolor[HTML]{D5E8D4}75.60}           & \cellcolor[HTML]{D5E8D4}2.36          & \cellcolor[HTML]{D5E8D4}423280                                                 & \multicolumn{1}{c|}{\cellcolor[HTML]{D5E8D4}108.184}         & \cellcolor[HTML]{D5E8D4}3.381          & \cellcolor[HTML]{D5E8D4}245.1 & \cellcolor[HTML]{D5E8D4}13.4                                                  \\
                                                                                          &                        & \cellcolor[HTML]{FFF2CC}5-Core (M)      & \multicolumn{1}{c|}{\cellcolor[HTML]{FFF2CC}\textbf{27.10}}  & \cellcolor[HTML]{FFF2CC}\textbf{0.85} & \cellcolor[HTML]{FFF2CC}\textbf{1180812}                                       & \multicolumn{1}{c|}{\cellcolor[HTML]{FFF2CC}40.542}          & \cellcolor[HTML]{FFF2CC}1.267          & \cellcolor[HTML]{FFF2CC}683.7 & \cellcolor[HTML]{FFF2CC}35.8                                                  \\ \cline{3-10} 
\multirow{-4}{*}{\begin{tabular}[c]{@{}c@{}}Human\\ Activity\\ ~\cite{human_activity_recognition_using_smartphones_240}\end{tabular}}         & \multirow{-4}{*}{84\%} & ESP32                                   & \multicolumn{1}{c|}{18528.00}                                & 579.00                                & 1727                                                                           & \multicolumn{1}{c|}{1451.113}                                & 45.347                                 & -                             & -                                                                             \\ \hline
                                                                                          &                        & \cellcolor[HTML]{CCCCFF}Base (B)        & \multicolumn{1}{c|}{\cellcolor[HTML]{CCCCFF}42.87}           & \cellcolor[HTML]{CCCCFF}1.34          & \cellcolor[HTML]{CCCCFF}746530                                                 & \multicolumn{1}{c|}{\cellcolor[HTML]{CCCCFF}\textbf{15.046}} & \cellcolor[HTML]{CCCCFF}\textbf{0.470} & \cellcolor[HTML]{CCCCFF}58.2  & \cellcolor[HTML]{CCCCFF}13.0                                                  \\
                                                                                          &                        & \cellcolor[HTML]{D5E8D4}Single Core (S) & \multicolumn{1}{c|}{\cellcolor[HTML]{D5E8D4}85.73}           & \cellcolor[HTML]{D5E8D4}2.68          & \cellcolor[HTML]{D5E8D4}373265                                                 & \multicolumn{1}{c|}{\cellcolor[HTML]{D5E8D4}122.680}         & \cellcolor[HTML]{D5E8D4}3.834          & \cellcolor[HTML]{D5E8D4}29.1  & \cellcolor[HTML]{D5E8D4}1.6                                                   \\
                                                                                          &                        & \cellcolor[HTML]{FFF2CC}5-Core (M)      & \multicolumn{1}{c|}{\cellcolor[HTML]{FFF2CC}\textbf{28.26}}  & \cellcolor[HTML]{FFF2CC}\textbf{0.88} & \cellcolor[HTML]{FFF2CC}\textbf{1132343}                                       & \multicolumn{1}{c|}{\cellcolor[HTML]{FFF2CC}42.277}          & \cellcolor[HTML]{FFF2CC}1.321          & \cellcolor[HTML]{FFF2CC}88.3  & \cellcolor[HTML]{FFF2CC}4.6                                                   \\ \cline{3-10} 
\multirow{-4}{*}{\begin{tabular}[c]{@{}c@{}}Gesture\\ Phase\\ ~\cite{gesture_phase_segmentation_302}\end{tabular}}          & \multirow{-4}{*}{89\%} & ESP32                                   & \multicolumn{1}{c|}{2496.00}                                 & 78.00                                 & 12821                                                                          & \multicolumn{1}{c|}{195.487}                                 & 6.109                                  & 1.0                           & 1.0                                                                           \\ \hline
                                                                                          &                        & \cellcolor[HTML]{CCCCFF}Base (B)        & \multicolumn{1}{c|}{\cellcolor[HTML]{CCCCFF}83.05}           & \cellcolor[HTML]{CCCCFF}2.60          & \cellcolor[HTML]{CCCCFF}385310                                                 & \multicolumn{1}{c|}{\cellcolor[HTML]{CCCCFF}\textbf{29.151}} & \cellcolor[HTML]{CCCCFF}\textbf{0.911} & \cellcolor[HTML]{CCCCFF}578.8 & \cellcolor[HTML]{CCCCFF}129.1                                                 \\
                                                                                          &                        & \cellcolor[HTML]{D5E8D4}Single Core (S) & \multicolumn{1}{c|}{\cellcolor[HTML]{D5E8D4}166.1}           & \cellcolor[HTML]{D5E8D4}5.19          & \cellcolor[HTML]{D5E8D4}192655                                                 & \multicolumn{1}{c|}{\cellcolor[HTML]{D5E8D4}237.689}         & \cellcolor[HTML]{D5E8D4}7.428          & \cellcolor[HTML]{D5E8D4}289.4 & \cellcolor[HTML]{D5E8D4}15.8                                                  \\
                                                                                          &                        & \cellcolor[HTML]{FFF2CC}5-Core (M)      & \multicolumn{1}{c|}{\cellcolor[HTML]{FFF2CC}\textbf{50.1}}   & \cellcolor[HTML]{FFF2CC}\textbf{1.57} & \cellcolor[HTML]{FFF2CC}\textbf{638723}                                        & \multicolumn{1}{c|}{\cellcolor[HTML]{FFF2CC}\textbf{74.950}} & \cellcolor[HTML]{FFF2CC}\textbf{2.342} & \cellcolor[HTML]{FFF2CC}959.4 & \cellcolor[HTML]{FFF2CC}50.2                                                  \\ \cline{3-10} 
\multirow{-4}{*}{\begin{tabular}[c]{@{}c@{}}Sensorless\\ Drives\\ ~\cite{dataset_for_sensorless_drive_diagnosis_325}\end{tabular}}      & \multirow{-4}{*}{86\%} & ESP32                                   & \multicolumn{1}{c|}{48,068.27}                               & 1502.13                               & 666                                                                            & \multicolumn{1}{c|}{3764.707}                                & 117.647                                & -                             & -                                                                             \\ \hline
                                                                                          &                        & \cellcolor[HTML]{CCCCFF}Base (B)        & \multicolumn{1}{c|}{\cellcolor[HTML]{CCCCFF}\textbf{30.115}} & \cellcolor[HTML]{CCCCFF}\textbf{0.94} & \cellcolor[HTML]{CCCCFF}\textbf{1062593}                                       & \multicolumn{1}{c|}{\cellcolor[HTML]{CCCCFF}\textbf{10.570}} & \cellcolor[HTML]{CCCCFF}\textbf{0.330} & \cellcolor[HTML]{CCCCFF}544.8 & \cellcolor[HTML]{CCCCFF}121.6                                                 \\
                                                                                          &                        & \cellcolor[HTML]{D5E8D4}Single Core (S) & \multicolumn{1}{c|}{\cellcolor[HTML]{D5E8D4}60.23}           & \cellcolor[HTML]{D5E8D4}1.88          & \cellcolor[HTML]{D5E8D4}531297                                                 & \multicolumn{1}{c|}{\cellcolor[HTML]{D5E8D4}86.189}          & \cellcolor[HTML]{D5E8D4}2.693          & \cellcolor[HTML]{D5E8D4}272.4 & \cellcolor[HTML]{D5E8D4}14.9                                                  \\
                                                                                          &                        & \cellcolor[HTML]{FFF2CC}5-Core (M)      & \multicolumn{1}{c|}{\cellcolor[HTML]{FFF2CC}57.57}           & \cellcolor[HTML]{FFF2CC}1.80          & \cellcolor[HTML]{FFF2CC}555845                                                 & \multicolumn{1}{c|}{\cellcolor[HTML]{FFF2CC}86.125}          & \cellcolor[HTML]{FFF2CC}2.691          & \cellcolor[HTML]{FFF2CC}285.0 & \cellcolor[HTML]{FFF2CC}14.9                                                  \\ \cline{3-10} 
\multirow{-4}{*}{\begin{tabular}[c]{@{}c@{}}Gas Sensor\\ Array Drift\\ ~\cite{gas_sensor_array_drift_dataset_224}\end{tabular}} & \multirow{-4}{*}{90\%} & ESP32                                   & \multicolumn{1}{c|}{16,407.33}                               & 512.73                                & 1950                                                                           & \multicolumn{1}{c|}{1285.022}                                & 40.157                                 & -                             & -                                                                             \\ \hline
\end{tabular}}
\vspace{-4.3mm}
\end{table*}

\textbf{Addressing Question 1: }Include-only \textit{encoding} based compression is just one way to compress the TM. Another Include-only approach for TMs has already been exploited in an FPGA inference implementation~\cite{MATADOR}. The FPGA approach (MATADOR) is seen in Fig.~\ref{fig:motivation}, it uses the Include TA actions to directly synthesize the model specific clause expressions in each class. The authors of MATADOR use Include-only sparsity to generate the closest comparable trade-off in terms of latency and performance per LUT. It uses the fewest LUTs of all comparable approaches and is the fastest of the existing TM accelerators~\cite{SveinConvTM}. However, it follows the design rationale of its contemporaries in Fig.~\ref{fig:motivation} as it creates model-specific accelerators. As such, the authors of FINN~\cite{FINN_R}, hls4ml~\cite{HLS4ML} and MATADOR offer open-source end-to-end automation flows to implement these designs, all requiring resynthesis every time. This work uses MATADOR's automation flow to replicate the TM model \textit{training} for MNIST, CIFAR 2~\cite{cifar} (2 classes: vehicles and animals) and Google Speech Commands~\cite{kws6} (Keyword Spotting with 6 words: yes, no, up, down, left, right - referred to as KWS 6) using the \textit{same} TM architectures, as mentioned in MATADOR, result in the same accuracy. The accelerator configurations are given in Table~\ref{tab:setup}. The proposed base \textbf{B} accelerator is the most resource efficient for LUTs and FFs and operates at the highest frequency. Additionally, on the same eFPGA chip, the Single Core (S) uses 2.5x fewer LUTs and 3.38x fewer FFs than MATADOR for MNIST. BRAMs for \textbf{B}, \textbf{S} and \textbf{M} are over-provisioned for more tunability later. The models were compressed into 16-bit instructions and used to program the proposed accelerator configurations. Fig.~\ref{fig:graph} shows the energy and latency of the configurations compared to MATADOR. The numbers in red indicate the speed-up and energy reduction compared to a software implementation of the compressed Include encoded inference approach running on an STM32Disco MCU (RDRS) as presented by~\cite{REDRESS} claiming upto 5700x speedup compared with embedded BNNs - MCU comparisons will be discussed in more detail in Question 2. All \textbf{B}, \textbf{S}, \textbf{M} configurations are within one order of magnitude of the MATADOR results; in the case of CIFAR 2, \textbf{B} is the \textit{most} energy efficient. The key point to note is that MATADOR is fixed to these latencies and energies; however, a real-time recalibration to a \textit{smaller} model would improve the \textbf{B}, \textbf{S}, \textbf{M} results without resynthesis. Ultimately, the key point becomes \textit{application}, if the trained model is always a good representation of the data it will infer, then MATADOR is a better option. If the data is subject to drift, or there is need for personalization or sensor reading degradation or environmental fluctuation, and there is an opportunity to use a system like Fig.~\ref{fig:recal}, then the proposed accelerators are more viable. They will adapt in situ during deployment.

\textbf{Addressing Question 2: }The flexibility to change model size and architecture at runtime makes the accelerator comparable to small RISC processors. Question 1 already demonstrated the energy and latency advantages against an Arm-based STM32Disco. This section explores how it compares to another even smaller, cheaper low-power MCU - the Espressif ESP32. This is done to understand whether the eFPGA is the best platform to be used in a system like Fig.~\ref{fig:recal}. Once again, the ESP32 runs the same compressed model inference, but as a software task on the processor. This time, the applications are chosen for their suitability for run-time recalibration (Table II). The EMG dataset involves classifying myographic signals sent from a bracelet to an edge inference node~\cite{emg_data_for_gestures_481}. This requires recalibration for user personalization, the same is true for Gesture Phase~\cite{gesture_phase_segmentation_302} and Human Activity detection~\cite{human_activity_recognition_using_smartphones_240}, they are all used to classify user movements. Sensorless Drives~\cite{dataset_for_sensorless_drive_diagnosis_325} involves diagnosing faulty components in electric current drive signals. Gas Sensor Array Drift uses chemical sensors to classify different gas concentrations~\cite{gas_sensor_array_drift_dataset_224}. Both applications are subject to environmental changes and component aging. Across all the datasets, all the proposed accelerators give better latency and energy efficiency (see speedup and energy reduction with respect to the ESP32 implementation). For Sensorless Drives the 5-Core \textbf{M} configuration gives the best energy performance. 

\vspace{-1mm}

\section{Conclusion}
\label{Sec:Conc}
This work explored the possibilities of LUT frugal and flexible accelerator development for applications that require runtime recalibration. The Tsetlin Machine algorithm leads to minimal bitwise compute, and its sparsity enables the instruction-based compression to allow models to fit well with the BRAMs of of-the-shelf eFPGA platforms. While current FPGA works leverage reconfigurability to build the fastest possible custom architectures, this work uses it to allow customization options in memory, batching, interconnect and number of cores. The proposed accelerator implementations offer far better energy efficiency than the low-power MCUs running the same compressed model inference.

\vspace{-1mm}
\section{Acknowledgments}
This work is supported by the UK Research and Innovation (UKRI) Engineering and Physical Sciences Research Council (EPSRC) under grant EP/X039943/1 and grant reference: studentship-2926263.

\bibliographystyle{ACM-Reference-Format}
\bibliography{acmart}

\end{document}